\documentclass[twocolumn,superscriptaddress]{revtex4}
\usepackage[pdftex,
 letterpaper=true,
 hyperindex=true,
 breaklinks=true,
 colorlinks=false,
 citecolor=blue,
 pdftitle={},
 pdfauthor={}]
{hyperref}
\usepackage{graphicx}
\usepackage{mathptmx}
\begin{document}

\title{Novel technique for constraining r-process (n,$\gamma$) reaction rates}

\author{A.~Spyrou}
    \email[]{spyrou@nscl.msu.edu}
	\affiliation{National Superconducting Cyclotron Laboratory, Michigan State University, East Lansing, Michigan 48824, USA}
	\affiliation{Department of Physics \& Astronomy, Michigan State University, East Lansing, Michigan 48824, USA}
	\affiliation{Joint Institute for Nuclear Astrophysics, Michigan State University, East Lansing, MI 48824, USA}

\author{S.~N.~Liddick}
    \email[]{liddick@nscl.msu.edu}
	\affiliation{National Superconducting Cyclotron Laboratory, Michigan State University, East Lansing, Michigan 48824, USA}
	\affiliation{Department of Chemistry, Michigan State University, East Lansing, Michigan 48824, USA}

\author{A.~C.~Larsen}
	    \email[]{a.c.larsen@fys.uio.no}
	\affiliation{Department of Physics, University of Oslo, NO-0316 Oslo, Norway}

\author{M.~Guttormsen}
	\affiliation{Department of Physics, University of Oslo, NO-0316 Oslo, Norway}

\author{K.~Cooper}
	\affiliation{National Superconducting Cyclotron Laboratory, Michigan State University, East Lansing, Michigan 48824, USA}
	\affiliation{Department of Chemistry, Michigan State University, East Lansing, Michigan 48824, USA}

\author{A.~C.~Dombos}
	\affiliation{National Superconducting Cyclotron Laboratory, Michigan State University, East Lansing, Michigan 48824, USA}
	\affiliation{Department of Physics \& Astronomy, Michigan State University, East Lansing, Michigan 48824, USA}
	\affiliation{Joint Institute for Nuclear Astrophysics, Michigan State University, East Lansing, MI 48824, USA}

\author{D.~J.~Morrissey}
	\affiliation{National Superconducting Cyclotron Laboratory, Michigan State University, East Lansing, Michigan 48824, USA}
	\affiliation{Department of Chemistry, Michigan State University, East Lansing, Michigan 48824, USA}

\author{F.~Naqvi}
 	\affiliation{National Superconducting Cyclotron Laboratory, Michigan State University, East Lansing, Michigan 48824, USA}

\author{G.~Perdikakis}
	\affiliation{Central Michigan University, Mt. Pleasant, Michigan, 48859, USA}
	\affiliation{National Superconducting Cyclotron Laboratory, Michigan State University, East Lansing, Michigan 48824, USA}
	\affiliation{Joint Institute for Nuclear Astrophysics, Michigan State University, East Lansing, MI 48824, USA}

\author{S.~J.~Quinn}
	\affiliation{National Superconducting Cyclotron Laboratory, Michigan State University, East Lansing, Michigan 48824, USA}
	\affiliation{Department of Physics \& Astronomy, Michigan State University, East Lansing, Michigan 48824, USA}
	\affiliation{Joint Institute for Nuclear Astrophysics, Michigan State University, East Lansing, MI 48824, USA}

\author{T.~Renstr{\o}m}
	\affiliation{Department of Physics, University of Oslo, NO-0316 Oslo, Norway}

\author{J.~A.~Rodriguez}
	\affiliation{National Superconducting Cyclotron Laboratory, Michigan State University, East Lansing, Michigan 48824, USA}

\author{A.~Simon}
	\affiliation{National Superconducting Cyclotron Laboratory, Michigan State University, East Lansing, Michigan 48824, USA}
	\affiliation{Department of Physics and The Joint Institute for Nuclear Astrophysics, University of Notre Dame, Notre Dame, Indiana 46556, USA}
	
\author{C.~S.~Sumithrarachchi}
  	\affiliation{National Superconducting Cyclotron Laboratory, Michigan State University, East Lansing, Michigan 48824, USA}

\author{R.~G.~T.~Zegers}
	\affiliation{National Superconducting Cyclotron Laboratory, Michigan State University, East Lansing, Michigan 48824, USA}
	\affiliation{Department of Physics \& Astronomy, Michigan State University, East Lansing, Michigan 48824, USA}
	\affiliation{Joint Institute for Nuclear Astrophysics, Michigan State University, East Lansing, MI 48824, USA}

\date{\today}

\begin{abstract}

A novel technique has been developed, which will open exciting new opportunities for studying the very neutron-rich nuclei involved in the r-process. As a proof-of-principle, the $\gamma$-spectra from the $\beta$-decay of $^{76}$Ga have been measured with the SuN detector at the National Superconducting Cyclotron Laboratory. The nuclear level density and $\gamma$-ray strength function are extracted and used as input to Hauser-Feshbach calculations. The present technique is shown to strongly constrain the $^{75}$Ge($n,\gamma$)$^{76}$Ge cross section and reaction rate.

 
\end{abstract}

\maketitle

One of the most important questions in Nuclear Astrophysics is the origin of the elements heavier than iron. It is well known that there are three main processes responsible for the nucleosynthesis of the heavier elements: two neutron-induced
processes (s- and r-process) that  create the majority of these nuclei and a third process (p-process), which is called upon to produce the small number of 
neutron-deficient isotopes that are not reached by the other two processes. Although the general characteristics of these processes were proposed already more than fifty years ago \cite{Bur57}, 
they are far from understood. 

Despite the fact that the r-process is responsible for producing roughly half of the isotopes of the heavy elements, its astrophysical
site has not yet been unambiguously identified. Multiple sites have been
proposed and investigated, however, to date, no firm conclusion has been drawn for where the r-process
takes place. Nevertheless, it is thought to occur in environments with a high density of free
neutrons, where neutron capture reactions push the matter flow to very neutron-rich nuclei, while
subsequent $\beta$-decays bring the flow back to the final stable nuclei (e.g. \cite{Arn07}). One of the limiting factors in
being able to determine the r-process site are the large uncertainties in the nuclear physics input. Because the nuclei 
involved in the r-process are many mass units away from the valley of stability, it is difficult, and sometimes even impossible
to measure the relevant quantities directly. A large effort has been devoted to the measurement of masses, $\beta$-decay half-lives, 
and $\beta$-delayed neutron emission probabilities (e.g. recently \cite{Hos10, Nis12, Van13}), however, the
majority of the r-process nuclei are still not accessible. In addition, although in many environments the neutron-capture reaction rates do not play significant role in the r-process flow due to ($n,\gamma$)-($\gamma,n$)
equilibrium, recent studies have shown significant sensitivity to the neutron-capture reaction rates in certain conditions \cite{Sur14}.
A major recognized challenge in the field is the measurement of the relevant neutron-capture reactions since all of the participating nuclei are unstable with short half-lives. The direct determination of the  ($n,\gamma$) cross sections that dominate in many cases the astrophysical r-process is not currently possible. It is therefore of paramount importance to develop indirect techniques to extract these critical reaction rates.

Many different techniques have been proposed for providing an indirect measurement of neutron-capture reaction rates far from stability, such as the surrogate reaction 
technique \cite{For07, Ciz07, Esc12}, 
 and the $\gamma$-ray strength function method ($\gamma$SF method) \cite{Uts10}. Significant effort is currently directed towards validating these techniques. In addition, very recently an idea for combining a radioactive ion beam facility with a reactor \cite{Rei14} has been proposed for direct measurement of ($n,\gamma$) reactions, although its application can take significant time and effort. 
 
 In this Letter we
introduce a novel technique for constraining neutron-capture reaction rates, which is based on the application of the well-known Oslo method \cite{Gut87, Sch00} combined with $\beta$-decay measurements using a $\gamma$-ray total absorption spectrometer (TAS). This technique provides an experimental determination of the nuclear level density (NLD) and the $\gamma$-ray strength function ($\gamma$SF), two quantities that together with the nucleon-nucleus optical model potential (OMP) define the neutron-capture cross section. The advantage of this technique is its applicability with very low beam
intensities (down to 1 particle per second or even lower), which allows one to reach farther from the valley of stability compared to the reaction-based techniques.

 The $\beta$-decay Q-values in neutron-rich nuclei increase systematically. As a result the study of NLD and $\gamma$SF can be done in a broad energy range, up to the neutron separation energy. The Oslo method is a proven technique and has been used extensively to extract NLD and $\gamma$SF along the valley of stability using various charged-particle reactions \cite{Gut12, Lar13}. In addition, it was shown that using these experimental NLD and $\gamma$SF as input for ($n,\gamma$) cross section calculations gives an excellent agreement with experimental cross section data \cite{Tor14}. The technique presented here offers
 a potential breakthrough in the 
measurement of these important nuclear properties far from stability and for extracting or, at the very least, constraining neutron-capture reaction rates along the r-process path.
In this Letter we demonstrate for the first time the application of this technique, hereafter called the  $\beta$-Oslo method, on the beta decay of $^{76}$Ga to constrain the reaction rate of the $^{75}$Ge(n,$\gamma$)$^{76}$Ge reaction. 

The experiment was performed at the National Superconducting Cyclotron Laboratory, at Michigan State University. A $^{76}$Ge primary beam was accelerated to the energy of 
130 MeV/u. A $^{76}$Ga secondary beam was produced from the fragmentation of the primary beam on a thick beryllium target. The $^{76}$Ga beam was stopped in the newly commissioned gas 
stopping area \cite{Coo14} and was extracted and delivered to the experiment with an intensity of $\approx$500 pps. No radioactive beam contaminants were observed after the gas stopping area.

The detection setup consisted of the Summing NaI (SuN) detector and a small silicon surface barrier detector. SuN is a $\gamma$-ray total absorption spectrometer that was recently developed
at the NSCL \cite{Sim13}. It is a cylindrical NaI(Tl) crystal, 16-inches in height and 16-inches in diameter, with a 1.8-inch in diameter bore-hole along its axis. SuN is segmented in 8 optically isolated segments, which can be used to observe individual $\gamma$ transitions. The full detector has a peak-efficiency of 85\% for the 661-keV $\gamma$-line of a $^{137}$Cs source. The signals from the eight segments can be summed to provide the total absorption spectrum, which is sensitive to the full energy available in a $\gamma$-cascade. During the experiment, a silicon surface barrier detector was placed at the center of SuN, and the $^{76}$Ga beam was implanted in that detector. Due to the low beam energy ($\approx$30 keV) the beam particles were stopped in the dead layer of the silicon detector and did not provide a measurable signal. After the decay of $^{76}$Ga ($T_{1/2}$= 32.6~s), the $\beta$ particles were detected in the silicon detector in coincidence with $\gamma$-rays in SuN. 

\begin{figure}[t]
 \begin{center}
 \includegraphics[clip,width=1.\columnwidth]{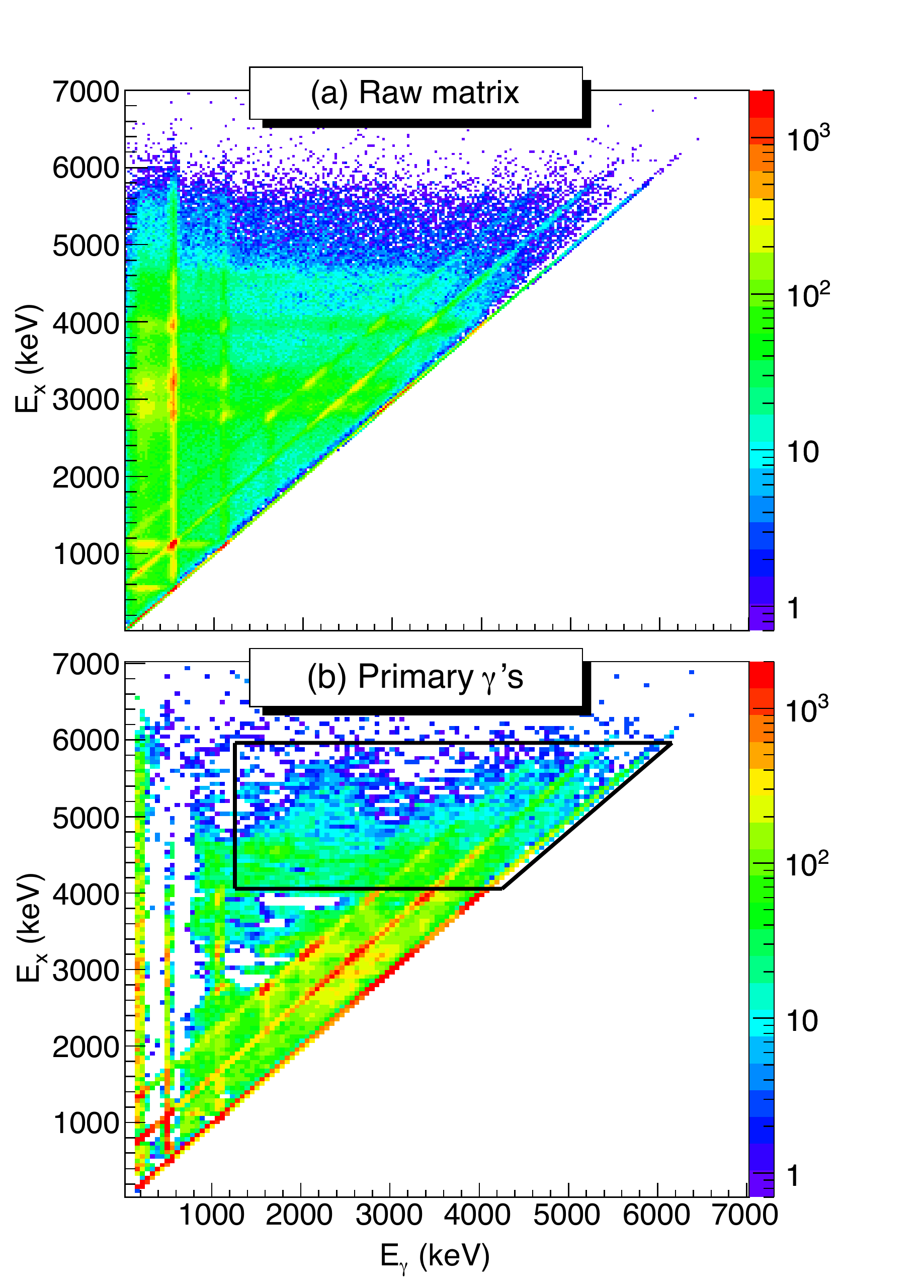}
 \caption {(Color online) (a) $^{76}$Ge ($E_\gamma,E_x$) matrix from $\beta^-$-decay of $^{76}$Ga:
 measured $\gamma$-ray energy in the NaI segments of the SuN detector versus the sum of all 
 $\gamma$-ray energies (total absorption spectrum), which equals the initial excitation energy $E_{x}$.
 The energy bins are 28 keV/channel. In total, the matrix has $\approx 860 000$ counts;
 (b) primary $\gamma$-ray distribution as a function of excitation energy. 
 The area within the solid, black lines is used for the extraction of level density
  and $\gamma$-ray strength function. The energy bins are 56 keV/channel.}
 \label{fig:matrix}
 \end{center}
 \end{figure}

To obtain information on the NLD and  $\gamma$SF of $^{76}$Ge, 
the Oslo method~\cite{Sch00} was applied. The raw coincidence ($E_\gamma,E_x$) matrix from the SuN detector
is shown in Fig.~\ref{fig:matrix}a. The excitation  energy $E_x$ was given by the total absorption spectrum, while the individual segments in SuN provided the $\gamma$-ray energy $E_\gamma$. 

The Oslo method relies on four main steps:
(\textit{i}) unfolding of the $\gamma$-spectra for each initial excitation energy~\cite{Gut96};
(\textit{ii}) isolation of the primary $\gamma$-ray spectrum, i.e. the distribution of the first emitted 
$\gamma$-rays in all the $\gamma$-decay cascades for each initial excitation energy~\cite{Gut87};
(\textit{iii}) extraction of the functional form of the level density and the $\gamma$-ray transmission coefficient
from the primary $\gamma$-ray spectra~\cite{Sch00};
(\textit{iv}) normalization of the NLD and $\gamma$SF~\cite{Sch00, Lar11}.

The unfolding of the $\gamma$-ray spectra was performed for each $E_x$ bin using the unfolding technique described in detail 
in Ref.~\cite{Gut96}, implementing the response functions for the segments of the SuN detector generated with 
the GEANT4 simulation tool~\cite{geant4a, geant4b}. The GEANT4 simulation for SuN was validated using standard radioactive sources and known resonances
as described in Ref.~\cite{Sim13}. The distribution of primary $\gamma$-rays was obtained through an iterative subtraction 
technique~\cite{Gut87}, where the primary $\gamma$-ray distribution for a given excitation-energy bin 
$E_j$ was determined by subtracting a weighted sum of the $\gamma$-spectra for all the underlying bins $E_{i<j}$. 
This technique has been thoroughly tested (see, e.g., Ref.~\cite{Lar11}), and is found to be reliable and robust
when the $\gamma$-decay routes from a given excitation-energy bin are the same regardless of how the states in 
that bin were populated (in this case, either directly via the $\beta$-decay of $^{76}$Ga or indirectly 
from $\gamma$-decay of higher-lying states). The matrix of primary $\gamma$-ray spectra for the full data set 
of $^{76}$Ge is shown in Fig.~\ref{fig:matrix}b. 

The primary $\gamma$-ray spectra represent the relative probability of a decay with $\gamma$-ray energy 
$E_\gamma$ from an
initial excitation-energy bin $E_x$, and depends on the level density at the final excitation energy 
$\rho(E_x - E_\gamma)$ and the $\gamma$-ray transmission coefficient $\mathcal{T}(E_\gamma)$~\cite{Sch00}:
\begin{equation}
P(E_\gamma,E_x) \propto \rho(E_x - E_\gamma) \cdot \mathcal{T}(E_\gamma),
\label{eq:ansatz}
\end{equation} 
where $P(E_\gamma,E_x)$ is the experimental primary $\gamma$-ray matrix. Using Eq.~(\ref{eq:ansatz}),
an iterative extraction procedure~\cite{Sch00} was applied to obtain the NLD and $\gamma$SF,
from the data within $E_{\gamma,\mathrm{min}} = 1.4$ MeV,
and $E_x \in [4.0,5.9]$ MeV (see Fig.~\ref{fig:matrix}b, and the Supplemental Material~\cite{suppl} for more details).

 \begin{figure}[t]
 \begin{center}
 \includegraphics[clip,width=1.\columnwidth]{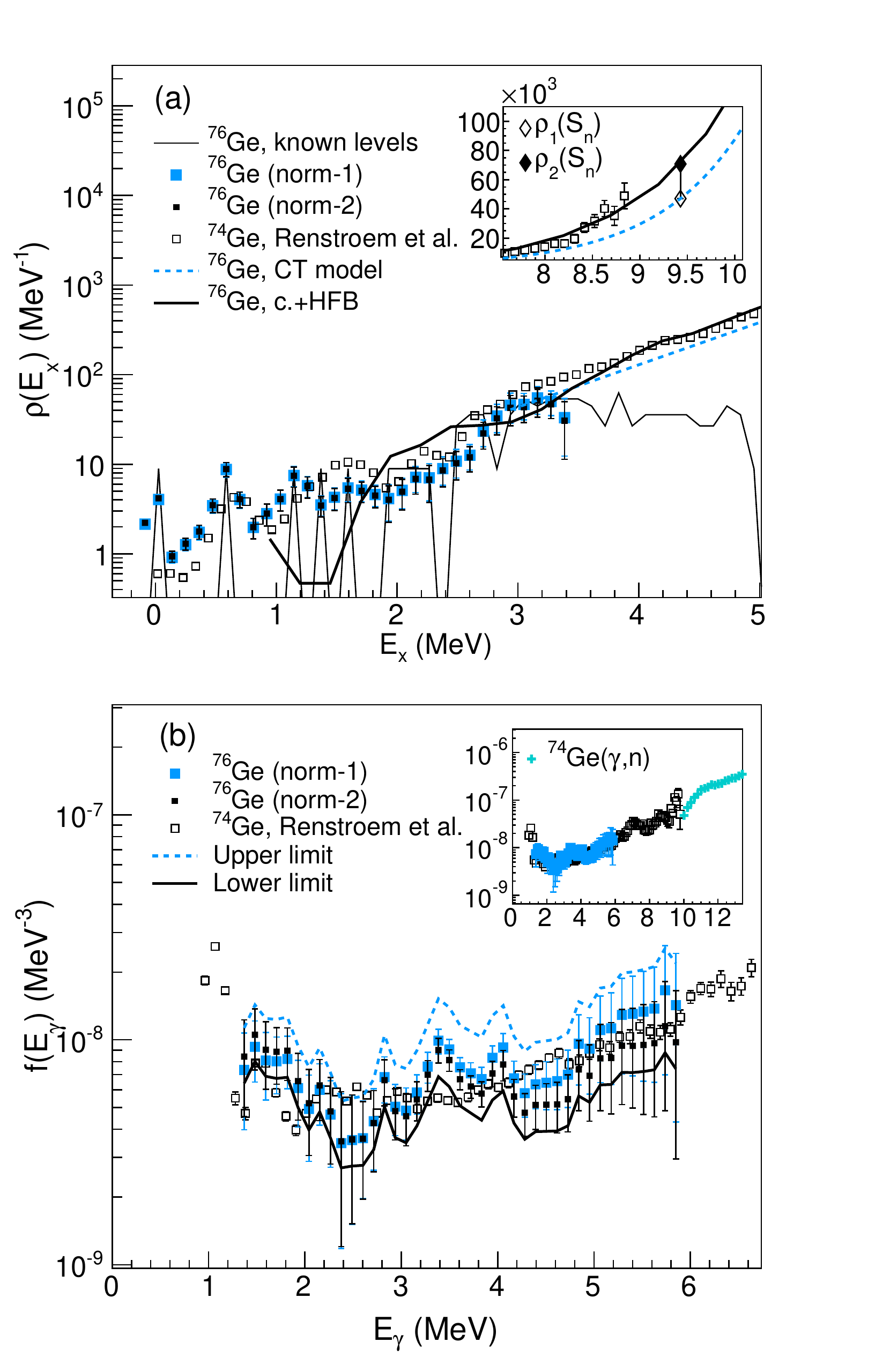}
 \caption {(Color online) (a) Level density of $^{76}$Ge compared to known, discrete levels, $^{74}$Ge data from Ref. \cite{Ren14}, the CT
model \cite{Egi09} with parameters $E_0 = -0.39$ MeV and $T = 0.92$ MeV, and the c.+HFB model \cite{Gor08} with an energy shift $\delta = -0.33$ MeV; the insert shows the models and the estimated 
 $\rho(S_n)$ for norm-1 and norm-2; 
 (b) $\gamma$SF of $^{76}$Ge for the different normalization procedures (see text for details), compared to $^{74}$Ge data \cite{Ren14}. The insert shows additional comparison with the $^{74}$Ge photo-neutron data from Ref. \cite{Gor82}.}
 \label{fig:nld_rsf}
 \end{center}
 \end{figure}

 \begin{figure*}[tb]
 \begin{center}
 \includegraphics[clip,width=2.\columnwidth]{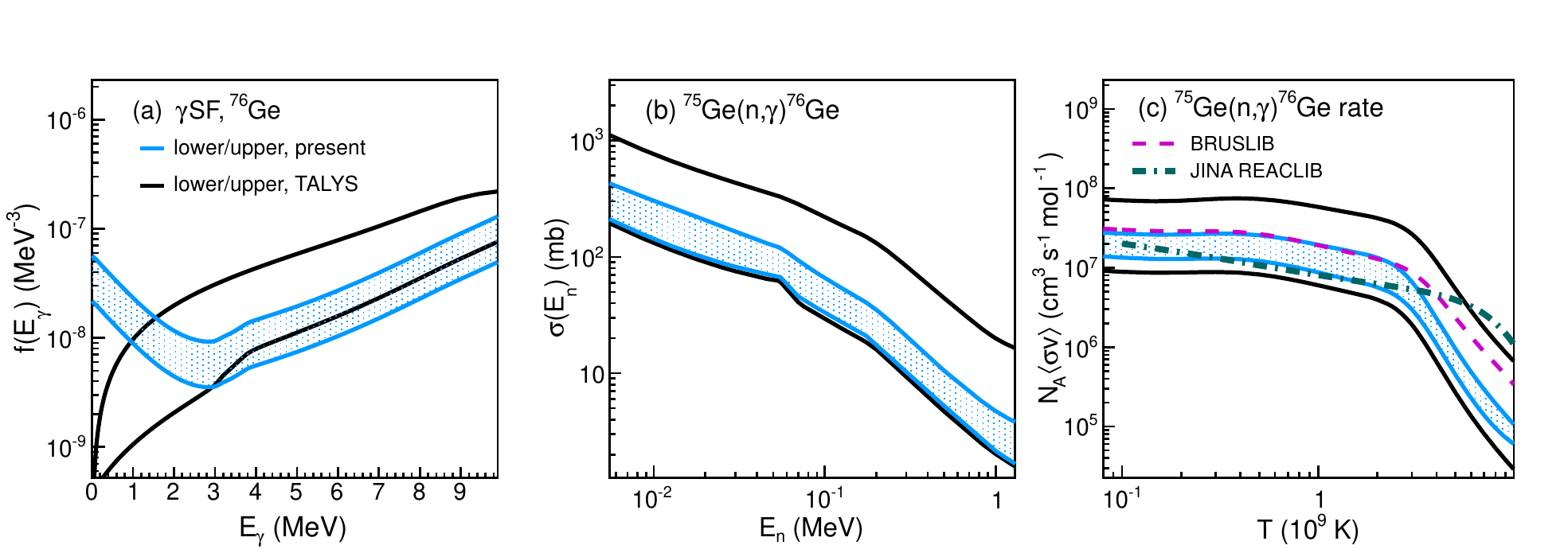}
 \caption {(Color online) The blue, filled area indicates the 
 constraints obtained with the present data, and the black lines indicate the lower and upper limits for the TALYS calculations 
 prior to the present work for 
 (a) the input $\gamma$SFs; (b) the $^{75}$Ge radiative neutron-capture cross section; (c) the Maxwellian-averaged reaction rates as function of  the stellar-environment temperature, also compared to rates from BRUSLIB~\cite{BRUSLIB} and JINA REACLIB~\cite{reaclib-v4}.
 }
 \label{fig:rates}
 \end{center}
 \end{figure*}

As only the functional form of the NLD and $\gamma$SF are obtained from the primary
$\gamma$-ray spectra, the slope and absolute value must be determined by other means.
Usually, known discrete levels at low $E_{x}$ and information from neutron-resonance experiments at the neutron separation energy 
$S_n$ have been used
for this purpose~\cite{Sch00}; however, no neutron-resonance data are available for $^{76}$Ge as $^{75}$Ge 
is unstable. Therefore, at $S_n$, the NLD was normalized to recent 
systematics~\cite{Egi09} using the
constant-temperature (CT) model~\cite{Eri59,Gil65} and was found to give an excellent description 
of available data~\cite{Voi09,Gut13}; hereafter, we refer to this normalization option as \textit{norm-1}.
This serves as a lower limit, as the spin distribution is rather narrow and centered at low spins.
Further, recent microscopic calculations based on the combinatorial-plus-Hartree-Fock-Bogoliubov (c.+HFB) 
approach using a Skyrme force~\cite{Gor08} have been applied, 
giving a significantly higher level density at $S_n$. This option is referred to 
as \textit{norm-2} and provides the upper limit, giving a broad spin distribution with a center-of-gravity at
rather high spins compared to norm-1. 
Thus, we have estimated for norm-1 (lower limit): $\rho_1(S_n) = 4.70\cdot 10^{4}$ MeV$^{-1}$, and for 
norm-2 (upper limit): $\rho_2(S_n) = 7.07\cdot 10^{4}$ MeV$^{-1}$. The normalized NLD of $^{76}$Ge is shown in
Fig.~\ref{fig:nld_rsf}a, and we observe an excellent agreement with the known, discrete levels. We also see that the
$^{76}$Ge data points resemble $^{74}$Ge data measured at the Oslo Cyclotron Laboratory~\cite{Ren14} as expected from previous studies of isotopic chains \cite{Mor14}. 
These findings give confidence in the present $\beta$-Oslo method.

Moreover, the $\gamma$SF is normalized to an average, total radiative width 
$\left< \Gamma_{\gamma0} \right> = 193^{+102}_{-46}$ meV estimated from systematics for the Ge isotopes, 
using neutron-resonance data 
from Ref.~\cite{Cap09}. 
The slope of the $\gamma$SF is deduced from a reduced value of $\rho(S_n)$ with the same
approach as recently applied for the actinides~\cite{Gut14}, as the 
$^{76}$Ga $\beta$-decay will populate levels with $J=1,2,3$ in $^{76}$Ge (the $^{76}$Ga ground-state spin is taken to be
2$^-$~\cite{Man11}). 
For further details on the normalization procedure 
and the applied parameters, see the Supplemental Material~\cite{suppl}. As the $\gamma$-decay in this excitation-energy region 
is dominated by dipole radiation~\cite{Kop90,Lar13}, the $\gamma$SF is deduced from the $\gamma$-transmission coefficient
by $f(E_\gamma) \simeq \mathcal{T}(E_\gamma)/2\pi E_\gamma^3$.

The normalized $\gamma$SF is shown in Fig.~\ref{fig:nld_rsf}b, where the error bars of the $^{76}$Ge data points include statistical 
errors, and propagated systematic errors from the unfolding and the primary $\gamma$-ray extraction. Additional systematic uncertainties
originating from the normalization process are indicated by the solid and dashed lines. Again, we find that the present data are
in overall very good agreement with the $^{74}$Ge data~\cite{Ren14}, as would be expected from previous observations, where neighboring isotopes display very similar $\gamma$SFs,  e.g. for Mo 
isotopes~\cite{Gut05}. We also see that the $^{76}$Ge $\gamma$SF is
increasing at low $\gamma$-ray energies. This \textit{upbend} phenomenon has been observed in many 
$fp$-shell~\cite{Lar13,Voi04,Lar06,Lar07,Voi10,Bur12} and $A\sim90-100$ nuclei~\cite{Gut05,Wie12},
and has the potential to significantly increase astrophysical ($n,\gamma$) reaction rates~\cite{Lar10} of paramount 
importance for the astrophysical r-process~\cite{Arn07}, in particular for conditions that are not under ($n,\gamma$)-($\gamma,n$) 
equilibrium \cite{Sur14}. Very recently, two theoretical explanations for this phenomenon have been published:
in the work of Ref.~\cite{Lit13} a low-energy increase appears in the $E1$ strength function as a result of thermal quasi-particle excitations into the continuum, while an enhanced $M1$ strength
is found in shell-model calculations of Ref.~\cite{Sch13} as a result of a re-orientation effect of high-$j$ neutron and proton
valence orbits. As of today, it is not known whether the observed behavior is due to either $E1$ or $M1$ transitions,
or both. It is therefore very interesting to study this phenomenon in unstable nuclei and map its strength far from stability. 

From the present analysis of the $^{76}$Ge data, the NLDs and $\gamma$SF were used as input in the TALYS-1.6 nuclear-reaction code~\cite{Kon08}, calculating the
($n,\gamma$) reaction cross section and Maxwellian-averaged reaction rate. Following Ref.~\cite{Sch13}, 
the $^{76}$Ge upbend was included as an $M1$ component
of the total dipole strength, with an exponential parametrization of the form $f_{\mathrm{up}}(E_\gamma) = C\exp{[-\eta E_\gamma]}$, with 
$C = 3.34 \times 10^{-8}$ MeV$^{-3}$ and $\eta=0.97$ MeV$^{-1}$. For the $E1$ $\gamma$-strength component, the Skyrme-HFB+QRPA calculation 
of Ref.~\cite{Gor02} was applied. In addition, the standard treatment of 
the $M1$ spin-flip resonance as described in the TALYS documentation is included~\cite{Kon08}. The total dipole strength is thus
$f(E_\gamma) = f_{\mathrm{up},M1}+f_{E1} + f_{\mathrm{spin-flip},M1}$.
For the experimental lower limit, we have used the CT model (norm-1) for the 
level density, $\left<\Gamma_{\gamma0}\right> = 147$ meV (scaling $f(E_\gamma)$ with a factor 0.65), and the
$JLM$ optical-model potential (JLM OMP)~\cite{Bau01,Kon08} (for more details see the Supplemental material \cite{suppl}).
For the experimental upper limit, the microscopic calculations of Ref.~\cite{Gor08} (norm-2) are applied, 
$\left<\Gamma_{\gamma0}\right> = 295$ meV (scaling $f(E_\gamma)$ with a factor 1.7), and using the 
neutron-optical-model potential (n-OMP) of Ref.~\cite{Kon03}. 

We have also tested the standard input options in TALYS to obtain the lower and upper limit as provided by TALYS, corresponding to:
(\textit{i}) a combinatorial-plus-HFB calculation with a Skyrme force~\cite{Gor08} for the level density,
the Skyrme-HFB+QRPA calculation 
of Ref.~\cite{Gor02}, and the
$JLM$ OMP~\cite{Bau01} (lower TALYS limit);
(\textit{ii}) the back-shifted Fermi-gas model as implemented in TALYS~\cite{Kon08}, the  
Brink-Axel model~\cite{Bri57,Axe62} for the E1 $\gamma$SF, and the n-OMP of Ref.~\cite{Kon03} (upper TALYS limit).
Note that the two OMP's are practically identical for incoming neutron energies between $\approx 50$keV$-$1 MeV, showing that the
uncertainties are dominated by the uncertainties in the NLD and $\gamma$SF. 

The results of our calculations are shown in Fig.~\ref{fig:rates} and the $(n,\gamma$) astrophysical reaction rate is also compared
to rates from the BRUSLIB~\cite{BRUSLIB} and from the JINA REACLIB~\cite{reaclib-v4}. We observe that our upper limit follows the 
BRUSLIB rate for temperatures below $\approx 2$ GK and our lower limit is in good agreement with the REACLIB rate. Both libraries overestimate the reaction rate at higher temperatures. 
We also note that despite the rather large uncertainties, we are able to significantly constrain
the ($n,\gamma$) cross section and the astrophysical ($n,\gamma$) reaction rate.  Hence, these results show that our new method has a great potential 
in further constraining astrophysical reaction rates for more neutron-rich nuclei, for which the $\beta$-decay $Q$-value will be comparable to
the neutron separation energy, and as such it could provide vital information both for fundamental nuclear structure and nuclear astrophysics.

In summary, the present Letter introduces a new technique that provides a unique opportunity for constraining ($n,\gamma$) cross sections far from stability. These cross sections are extremely important for the astrophysical r-process and currently the tools for studying these reactions are at best limited. 
The presented method combines the use of $\beta$-decays to populate high-lying levels in the nucleus of interest with a segmented total absorption spectrometer for detecting the individual $\gamma$ rays and excitation energy and with the well known Oslo method for extracting nuclear level densities and $\gamma$-ray strength functions. Employing the $\beta$-decay as a means to populate the levels of interest greatly increases the number of nuclei within experimental reach and allows in many cases to reach the r-process path at current and next generation facilities. 

The authors gratefully acknowledge the support of NSCL operations staff. 
A.~C.~L. and M.~G. gratefully acknowledge financial support from the Research Council of Norway, 
project grant no. 205528. This work was supported by the National
Science Foundation under Grants No. PHY~102511 (NSCL) and
No. PHY~0822648 (Joint Institute for Nuclear Astrophysics), and PHY~1350234 (CAREER).




\end{document}